\documentclass[sigconf]{acmart}

\usepackage{titlesec}

\usepackage{enumitem}
\usepackage{multirow}
\usepackage{listings}
\usepackage{sparklines}
\usepackage[frozencache,cachedir=minted-cache]{minted}
\usepackage{listings}
\usepackage[most]{tcolorbox} %

\newcommand{\TODO}[1]{\textcolor{red}{#1}\GenericWarning{}{LaTeX Warning: TODO:#1}}\newcommand\todo\TODO
\usepackage{xifthen}
\newcommand{\newc}[2]{%
  \ifstrempty{#1}{%
    \ifstrempty{#2}{}{%
      \textcolor{black}{#2}%
    }%
  }{%
    \ifstrempty{#2}{%
      \textcolor{black}{#1}%
    }{%
      \textcolor{red}{#1} $\rightarrow$ \textcolor{black}{#2}%
    }%
  }%
}

\usepackage[most]{tcolorbox}
\usepackage{cleveref}
\AtBeginDocument{%
  }
\Crefname{figure}{Fig.}{Fig.}

\definecolor{dark_orange}{HTML}{F66F4D}
\definecolor{light_orange}{HTML}{f8be92}
\definecolor{turquesa}{HTML}{a5d4cc}
\definecolor{light_blue}{HTML}{B5CEDA}
\usepackage{transparent}

\lstdefinelanguage{yaml}{
  keywords={true,false,null,y,n,yes,no},
  keywordstyle=\color{blue}\bfseries,
  keywords={name},
  keywordstyle=\color{turquesa},
  sensitive=false,
  comment=[l]{\#},
  morecomment=[s]{/*}{*/},
  morestring=[b]",
  morestring=[b]',
  stringstyle=\color{orange},
  commentstyle=\color{gray}
}

\tcbset{
    action_style/.style={
        colback = light_blue,  %
        colframe = orange!20!black, %
        boxrule = 0.8pt,         %
        arc = 3pt,               %
        left = 3pt, right = 3pt, %
        top = 3pt, bottom = 3pt
    }
}
\setcopyright{acmlicensed}
\copyrightyear{2026}
\acmYear{2026}
\acmDOI{XXXXXXX.XXXXXXX}
\acmConference[EASE 2026]{The 30th International Conference on Evaluation and Assessment in Software Engineering}{9–12 June, 2026}{Glasgow, Scotland, United Kingdom}
\acmISBN{978-1-4503-XXXX-X/2018/06}

\begin{document}

\title{The Ultimate Configuration Management Tool? \\Lessons from a Mixed Methods Study of Ansible’s Challenges}

\author{Carolina Carreira}
\authornote{Both authors contributed equally to this research.}
\email{carolina.carreira@tecnico.ulisboa.pt}
\affiliation{%
  \institution{Carnegie Mellon University, INESC-ID, and \\IST, University of Lisbon}
  \city{Lisbon}
  \country{Portugal}
}

\author{Nuno Saavedra}
\authornotemark[1]
\email{nuno.saavedra@tecnico.ulisboa.pt}
\affiliation{%
  \institution{INESC-ID and IST, University of Lisbon}
  \city{Lisbon}
  \country{Portugal}
}

\author{Alexandra Mendes}
\email{alexandra@archimendes.com}
\affiliation{%
  \institution{Faculdade de Engenharia,\\
Universidade do Porto}
\city{Porto}
  \country{Portugal}
}

\author{João F. Ferreira}
\email{joao@joaoff.com}
\affiliation{%
  \institution{INESC-ID and Faculdade de Engenharia,\\ Universidade do Porto}
  \city{Porto}
  \country{Portugal}
}

\renewcommand{\shortauthors}{Carreira et al.}

\begin{abstract}

Infrastructure as Code (IaC) tools have transformed the way IT infrastructure is automated and managed, but their growing adoption has also exposed numerous challenges for practitioners. In this paper, we investigate these challenges through the lens of Ansible, a popular IaC tool. %
Using a mixed methods approach, we investigate challenges faced by practitioners. We analyze 59,157 posts from Stack Overflow, Reddit, and the Ansible Forum to identify common pain points, complemented by 20 semi-structured interviews with practitioners of varying expertise levels.

Based on our findings, we highlight key directions for improving Ansible, with implications for other IaC technologies, including stronger failure locality to support debugging, clearer separation of language and templating boundaries, targeted documentation, and improved execution backends to address performance issues. By grounding these insights in the real-world struggles of Ansible users, this study provides actionable guidance for tool designers and for the broader IaC community, and contributes to a deeper understanding of the trade-offs inherent in IaC tools.

\end{abstract}


\hyphenation{DevOps}

\definecolor{blueturquoise}{RGB}{51, 204, 204} %

\definecolor{darkgreen}{RGB}{8,144,0}

\newcommand{\icircled}[1]{\tikz[baseline=(char.base)]{
            \node[shape=circle,draw=blueturquoise,text=blueturquoise,inner sep=0.5pt,scale=0.85] (char) {#1};}}
\newcommand{\tcircled}[1]{\tikz[baseline=(char.base)]{
            \node[shape=circle,draw=darkgreen,text=darkgreen,inner sep=0.5pt,scale=0.85] (char) {#1};}}

\maketitle

\section{Introduction}
\label{sec:intro}

Infrastructure as Code (IaC) has emerged as a critical paradigm in modern software engineering, enabling organizations to automate the configuration, provisioning, and management of IT infrastructure. By treating infrastructure configurations as code, IaC enables practitioners to use software engineering practices such as automated testing and version control, and facilitates reproducibility, scalability, and maintainability of infrastructure resources.

IaC has become widespread across industries, with several tools gaining significant adoption. Examples of widely used IaC tools include Ansible~\cite{ansible}, Chef~\cite{chef}, and Puppet~\cite{puppet} for configuration management, and Terraform~\cite{terraform} and CloudFormation~\cite{cloudformation} for infrastructure provisioning. These tools allow organizations to automate complex infrastructure tasks, reducing manual effort and the risk of human error. For example, NASA and Siemens use Ansible to automate infrastructure deployment and configuration management~\cite{redhat_siemens,nasa_signal}. The widespread adoption of these tools highlights their importance in modern software operations.

Despite the significant advantages of IaC, practitioners face several challenges. Since IaC involves writing code, errors can be introduced, leading to misconfigurations that affect system security~\cite{rahman2019seven,rahman2021security,reis2022leveraging,saavedra2022glitch,saavedra2023polyglot,opdebeeck2023control} and system reliability ~\cite{gitlab_outage_2014,stack_exchange_outage_2014,reddit_outage_2016,incident_resolution_honeycomb_2021,rust_outage_2023, wikimedia_dataloss_2016}. For example, due to bugs in their IaC scripts, GitHub experienced an outage of its DNS infrastructure~\cite{fryman_2014} and Amazon Web Services lost around 150 million dollars after issues with its S3 billing system~\cite{hersher_2017}. To address this, there has been an effort by the research community to categorize, identify and repair defects~\cite{saavedra2022glitch,saavedra2023polyglot,rahman2021security,reis2022leveraging,hassan2024state,begoug2024fine,sotiropoulos2020practical,opdebeeck2022smelly,rahman2023security,lepiller2021analyzing,schwarz2018code,sharma2016does,bessghaier2024prevalence,opdebeeck2023control,zerouali2023helm,rahman2018characterizing,rahman2019seven,rahman2020gang,saavedra2025infrafix}. In addition, IaC tools themselves present challenges, such as unintuitive debugging mechanisms, inconsistent behaviors, and complex learning curves. Prior research has explored challenges faced by IaC practitioners, identifying common issues and limitations of existing tools~\cite{rahman2018questions,guerriero2019adoption,begoug2023infrastructure,tanzil2023mixed}. However, existing studies often take a broad perspective, examining challenges across multiple DevOps or IaC tools, which can dilute insights specific to any one tool. Studies focused on a single IaC tool have not, so far, included Ansible. %

In this paper, we aim to deepen the understanding of the challenges faced by Ansible users. 
Ansible is currently the most widely used configuration management tool~\cite{Overflow_2024} and, although Ansible is known to have issues, such as difficulties with variable scoping~\cite{opdebeeck2024static} and debugging~\cite{hackernews_ansible_2021}, the specifics of these challenges and their impact on practitioners remain underexplored. Using a mixed methods approach, we analyze a large corpus of online discussions and conduct semi-structured interviews to systematically examine the obstacles faced by Ansible users. Our study is the first to provide a detailed account of these issues and to offer actionable recommendations for improvement. %
Our contributions are: %
\begin{itemize}[leftmargin=*]
\item A large-scale empirical analysis of 59,157 online discussions to identify common challenges and issues in Ansible.

\item Insights from 20 semi-structured interviews with Ansible practitioners of varying expertise levels, providing qualitative perspectives on usability and technical challenges.

\item Recommendations for improving Ansible, with implications for other IaC technologies, focusing on stronger failure locality to support debugging, better defined language boundaries, targeted documentation, and improved execution backends.

\end{itemize}
All supplementary materials, data, and code are available in a replication package~\cite{replicationpackage}.

\section{Ansible}
\label{sec:background}

Ansible is the most widely used configuration management tool~\cite{Overflow_2024}.
Configurations are written as YAML playbooks composed of plays and tasks, optionally organized into reusable roles.
Tasks call modules with arguments to enforce the desired state on target hosts.
For example, the \emph{ansible.builtin.shell} module runs shell commands on the host. 
Ansible executes tasks sequentially in the order in which they appear in a playbook. It uses variables to manage system differences, allowing the same variable name to hold different values for each host. These variables can be used in playbooks to adjust task behavior based on the target host. 
Ansible leverages Jinja2 templating to enable dynamic expressions and access to variables~\cite{ansible_docs}, both within the playbooks and in configuration files used via the \emph{template} module.
Unlike pull-based technologies such as Chef or Puppet, Ansible follows an agentless, push-based model, typically applying changes from a control node to target hosts over SSH.

\Cref{fig:ansible-example} shows excerpts of an Ansible playbook for the configuration of a web server, consisting of a single play with three tasks. The playbook targets all hosts in the inventory, uses Ansible variables and Jinja2 expressions for conditional execution, and applies a template to generate a customized web page.

\begin{figure}
    \centering
\begin{minted}[linenos=true,fontsize=\scriptsize,xleftmargin=2em, escapeinside=!!]{yaml}
---
- name: Set up and configure web server
  hosts: all !\label{line:hosts}!
  gather_facts: yes
  tasks:
    - name: Display system facts
      debug:
        msg: "The system OS is {{ ansible_facts['os_family'] }}." !\label{line:fact}!
      when: ansible_facts['os_family'] == 'Debian' !\label{line:condition}!
    - name: Ensure Nginx is installed
      apt:
        name: nginx
    - name: Deploy an index.html template
      template: !\label{line:template}!
        src: templates/index.html.j2
        dest: /var/www/html/index.html
    (...)
\end{minted}
    \caption{Excerpt from a web server Ansible playbook. %
    }
    \label{fig:ansible-example}
\end{figure}

\section{Mixed Methods Study}

\begin{figure*}
    \centering
    \includegraphics[width=0.9\linewidth]{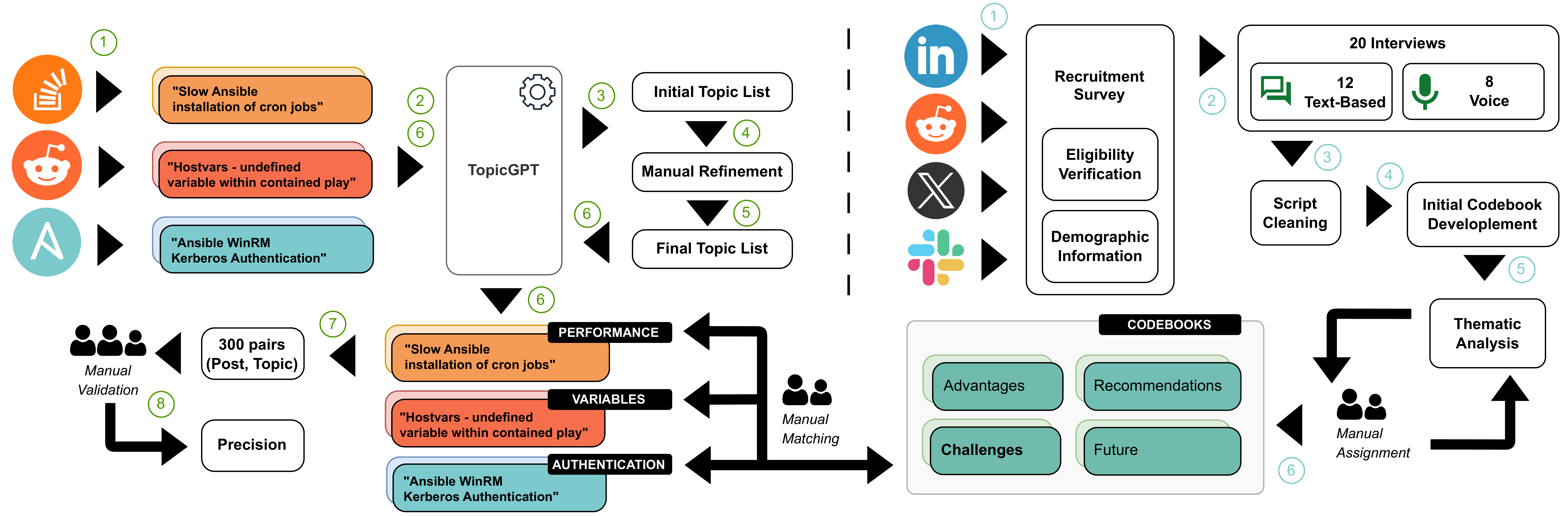}
    \caption{Overview of the research methodology combining automated topic extraction and thematic analysis. %
    }
    \label{fig:overview}
\end{figure*}

In this section, we outline our research questions and describe the methodology used to answer them. Our study follows a mixed methods approach (see \Cref{fig:overview}). First, we collected posts from Q\&A platforms and used topic modeling to understand the challenges faced by Ansible practitioners. We then conducted semi-structured interviews to gain deeper insights into the participants' experiences. %

\subsection{Research Questions}
\label{sec:research-questions}
We aim to answer the following research questions.

\begin{itemize}[label={},leftmargin=*]
  \item \textbf{RQ1: [Issues]} What are the common issues practitioners face when using Ansible?
  In this question, we investigate the common challenges that practitioners face. %
  We analyze topics from practitioners' posts in Q\&A platforms and feedback from interviews to categorize these challenges and identify patterns that can help us to understand pain points across various praticioner groups.
  \item \textbf{RQ2: [Adoption]} What are the aspects that influence the adoption of Ansible? 
  This question focuses on the factors that drive Ansible's adoption, identified via interviews with practitioners. 
  \item \textbf{RQ3: [Improvements]} What changes would improve Ansible?
  This question aims to identify actionable suggestions for improving Ansible. The goal is to provide insights for Ansible's development community, and broader IaC community, on how to refine the tool to better meet practitioners' expectations and needs.   
\end{itemize}

RQ1 is addressed using both topics extracted from online Q\&A data and practitioner interviews, enabling triangulation between large-scale community discussions and in-depth qualitative insights. RQ2 and RQ3 are answered exclusively through interviews, as public technical artifacts primarily reflect post-adoption troubleshooting and rarely capture adoption rationales or broader improvement considerations. The key directions discussed in Section~\ref{sec:discussion} are grounded in the combined results of all three research questions, integrating evidence from both data sources.

\subsection{Topic Modeling}
\label{sec:topic_model}

Building on previous studies \cite{begoug2023infrastructure,tanzil2023mixed}, we apply topic modeling techniques to identify the challenges of using Ansible. We use TopicGPT, a prompt-based framework
that uses LLMs to uncover latent topics in a text collection~\cite{pham2024topicgpt}. TopicGPT provides interpretable topics that align better with human categorizations than other topic modeling approaches~\cite{pham2024topicgpt}. After data is collected, TopicGPT consists of three steps: \emph{Topic Generation}, \emph{Topic Refinement}, and \emph{Topic Assignment}. We also perform manual validation of the \emph{Topic Assignment} step. We configured TopicGPT to use GPT-4o mini via the OpenAI API. %
\newc{}{We used TopicGPT’s original prompts, adapting them to our task by replacing generic examples with Ansible-specific ones and illustrative post assignments. We also refined the prompts to ensure the topics captured challenges faced by Ansible practitioners and that topic assignments reflected this focus.}

\subsubsection*{Data Collection}
We start by collecting posts about challenges, obstacles, and issues faced by Ansible practitioners from three Q\&A platforms: StackOverflow~\cite{stackoverflow}, Reddit~\cite{reddit}, and the Ansible Forum~\cite{ansibleforum} (\tcircled{1} in~\Cref{fig:overview}). To collect posts from StackOverflow, we used the Stack Exchange API~\cite{stackexchange} and retrieved all posts with the tag \emph{Ansible}, which corresponds to 23,000 posts. For Reddit, we used the PullPush API~\cite{pullpush} to collect all 7,187 posts on the subreddit \emph{r/ansible} that contained the keyword \emph{Ansible} in addition to keywords possibly related to issues, such as \emph{error} and \emph{help}. For the Ansible Forum, we used web scraping techniques to collect 28,970 posts in the \emph{Get Help} and \emph{Archives} categories. We included the category \emph{Archives} in our analysis, as it contains all posts from the \emph{Ansible Google Group}, the predecessor of the Ansible Forum, which were migrated to the Ansible Forum~\cite{ansibleforum_move}. We filter out posts related to announcements by the Ansible team. In total, we collected 59,157 posts from the three Q\&A platforms, forming our final dataset. Data collection took place between January 24 and February 1, 2025.

\subsubsection*{Topic Generation}

We use TopicGPT to generate a set of topics given our whole dataset as input (\tcircled{2} in~\Cref{fig:overview}). A topic consists of a name and a description. For each post, TopicGPT prompts GPT-4o Mini to assign one or more of the already generated topics and generate any new relevant topics that are missing, assigning them accordingly. 
To ensure relevance, we constrained the TopicGPT prompt to generate topics centered on the challenges faced by Ansible practitioners.
In the end, TopicGPT generated 3,573 topics (\tcircled{3} in~\Cref{fig:overview}).

\subsubsection*{Topic Refinement}
\label{sec:topic_refinement}

TopicGPT produces numerous topics that, while generally relevant, are often overly granular, duplicated, or out of scope \cite{pham2024topicgpt}. 
In our preliminary experiments, the automated refinement provided by TopicGPT did not reliably merge semantically equivalent topics.
For this reason, we followed a TopicGPT-like approach but used manual effort instead of LLMs to ensure a reliable final topic list (\tcircled{4} in~\Cref{fig:overview}).
Since topic frequencies follow a power-law distribution, we retained the most frequent topics until they covered 80\% of all attributions, discarding the rest. This reduced the topic list from 3,573 to 153.

Next, we applied an iterative manual refinement similar to TopicGPT’s merging step. In each iteration, topic pairs above a cosine similarity threshold were evaluated: one author decided whether to merge them based on semantic overlap and whether to assign them to an existing or new topic, while a second author reviewed the decisions and flagged disagreements. Both authors then resolved discrepancies through discussion. After six iterations, no further merges were identified, reducing the topic list from 153 to 105.

Finally, following prior work~\cite{asmussen2019smart,jacobi2018quantitative}, we manually inspected the topics to identify those relevant to our study. A topic was considered relevant if it is related to the challenges, frustrations, issues, or drawbacks associated with using Ansible. We considered only topics that were specific enough to provide meaningful information about a post. 
We also refined topic names and descriptions for clarity and generality and merged any remaining topic pairs that met our criteria, but were not identified earlier.

An author reviewed each topic to decide whether it should be removed or modified, while a second author independently reviewed the changes and flagged disagreements. Both authors then resolved the discrepancies through discussion, resulting in a final list of 87 topics (\tcircled{5} in~\Cref{fig:overview}; see Supplemental Materials for the full list).

\subsubsection*{Topic Assignment}
\label{sec:topic_assignment} 

We used TopicGPT to assign the final topics to posts and applying its self-correction mechanism~\cite{pham2024topicgpt} (\tcircled{6} in~\Cref{fig:overview}).

\subsubsection*{Validation}

We evaluated topic assignment using confirmation labeling~\cite{yang2014large}.
To calculate the precision of topic assignment, this method involves sampling \emph{(post, topic)} pairs of assignments under evaluation. Rather than presenting raters with the entire taxonomy and asking them to manually assign topics to posts, confirmation labeling simplifies the process by requiring raters to provide binary judgments on whether a given topic is relevant to a particular post. This approach effectively reduces the cognitive overload of a large and complex taxonomy~\cite{yang2014large}. 
Two authors independently performed confirmation labeling on a sample of 300 \emph{(post, topic)} pairs (\tcircled{7} in~\Cref{fig:overview}). 
A third author resolved disagreements between the initial raters, with final decisions made by majority vote, yielding a precision of 82.67\% (\tcircled{8} in~\Cref{fig:overview}).

\subsubsection*{Analysis}\label{subsec:topic_analysis} 
To integrate topic modeling and interview findings, we aligned the final topics with the interview \emph{Challenges} codes from Section~\ref{sec:interviews}. This process was iterative and followed a double-blind procedure: first, two coders independently grouped the topics into themes. Then, both coders met to discuss disagreements and reach a consensus.
In analyzing the merged data, we examined the relative frequency of each topic across the resulting themes and investigated how these frequencies evolved. \textbf{To better reflect current and emerging concerns, we limited our analysis to Q\&A data from the five years preceding the end of data collection on February 1, 2025, resulting in a total of 24,132 posts.} In addition to recent posts, we leveraged the data older than five years to identify evolution trends.

\subsection{Interviews}
\label{sec:interviews}

In addition to the large-scale analysis of Q\&A posts, we conducted semi-structured interviews to gain deeper insight into practitioners' experiences with Ansible. %

\subsubsection*{Participants}
Participants were categorized by Ansible expertise using years of experience, following prior work~\cite{latoza2010developers, roehm2012professional}, and to mitigate self-assessment bias, such as the Dunning–Kruger effect~\cite{mahmood2016people}. We define three levels: \textbf{Beginner} ($<2$ years, n=5), \textbf{Intermediate} ($2\geq$ and  $\leq5$ years, n=7), and \textbf{Experienced} ($>5$ years, n=8).

\subsubsection*{Recruitment}
We recruited 20 participants through a pre-screening survey distributed via email and social media (e.g., LinkedIn, X, Discord, and Slack (\icircled{1} in~\Cref{fig:overview})). To meet our eligibility criteria, participants had to have prior experience with Ansible. 
Participants could select their preferred interview modality: text or video. The availability of both formats was intended to address known challenges in recruiting software developers for research studies~\cite{baltes2016worse}. 
Of 127 survey participants, 118 were eligible, and 20 were available for an interview (15.7\%). Most interviewees (12/20) chose text, with several saying they could only participate in that format.
As an incentive, interviewees could enter a raffle for a \$100 gift card. Demographic details appear in the Supplemental Materials. Participants reported 1 to 9+ years of Ansible experience, most were active users (15), male (17), aged 18--34 (11), held at least a bachelor's degree (17), and had CS education and work experience (17).

\subsubsection*{Ethical Considerations}
All participants reviewed the consent form and provided informed consent. %
This study was approved by the Ethics Committee of Instituto Superior Técnico (Ref. 23/2024 (CE-IST)).

\subsubsection*{Interviews}

We conducted semi-structured interviews to obtain a nuanced understanding of the participants' experiences with Ansible (\icircled{2} in~\Cref{fig:overview}). Semi-structured interviews are the most widely used interviewing method~\cite{dicicco2006qualitative} and allow adaptability to explore emerging themes and seek clarifications~\cite{lazar2017research}. We piloted the protocol with four participants. %
Both text and voice-based interviews required around 30 minutes of attention time. The latter were recorded and automatically transcribed with participant consent using Zoom's built-in transcription functionality. Research team members manually reviewed the resulting transcripts. %
\newc{}{The interview protocol was developed independently of the topic modeling results.}
The interview began with background questions, then covered Ansible's \textit{advantages}, adoption \textit{challenges}, \textit{recommendations} for improvement, and \textit{future} perspectives. We sought clarification whenever responses were ambiguous or needed elaboration in both text and voice interviews.

\subsubsection*{Analysis}\label{subsubsec:analysiscoding}    
We developed four emergent codebooks using thematic analysis~\cite{braun2006using, lazar2017research}, following a bottom-up approach that allowed themes and codes to emerge inductively from our data. %
We started by reading the transcripts to highlight recurring ideas, creating initial codes, and iteratively refining them into coherent themes (\icircled{3} and \icircled{4} in~\Cref{fig:overview}).
Our codebooks are related to the four key thematic areas as follows:
(1) \textbf{Advantages}: Identifying the key motivations behind adoption;
(2) \textbf{Challenges}: Documenting difficulties and limitations; %
(3) \textbf{Recommendations}: Capturing participants' wishes; %
(4) \textbf{Future}: Addressing potential future developments. %

For each of the four areas, we developed a codebook following prior work~\cite{10548458,10548562,karlsson2007requirements, lazar2017research}. %
One author drafted the initial codebook, which a second author revised using a subset of interviews. We repeated this until the codebook stabilized. We then double-coded all interviews.%
The coders met at the end to resolve discrepancies and reach a consensus (\icircled{5} and \icircled{6} in~\Cref{fig:overview}).
Quotes are labeled by participant ID and experience level: (B)eginner, (I)ntermediate, or (E)xperienced (e.g., B1 denotes a beginner). The full codebooks are in the Supplemental Materials.

\label{sec:methodology}

\section{RQ1: Common Issues in Ansible Usage}\label{sec:RQ1}
\begin{figure*}
    \centering
    \includegraphics[width=1\linewidth]{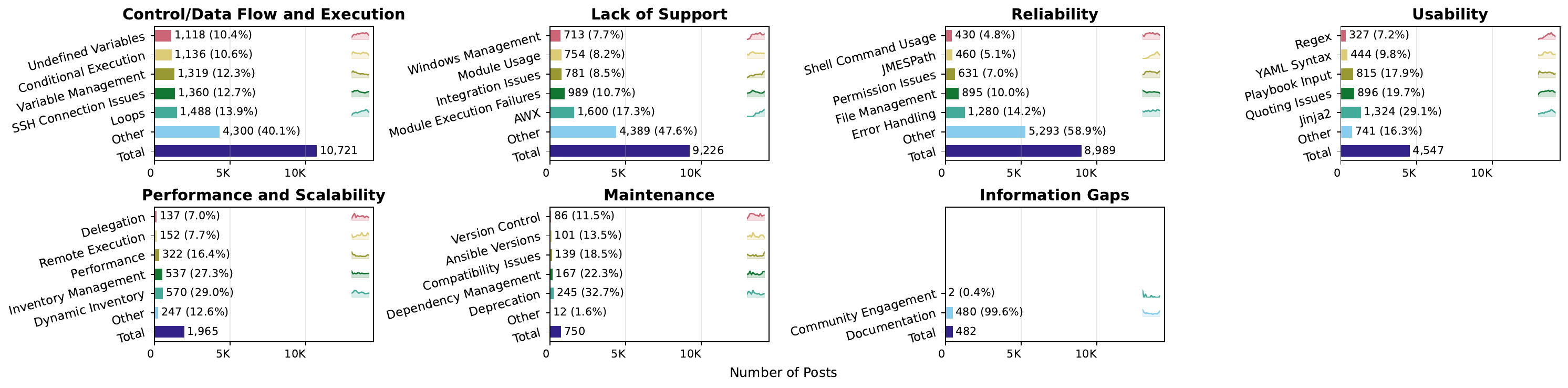}
    \caption{Topic distribution (Feb 2020–Feb 2025). Sparklines show yearly percentage trends across the full Ansible post history. %
    }
    \label{fig:topic-modeling-challenge-categories}
\end{figure*}

This section presents the seven primary challenge categories encountered by Ansible practitioners, ordered by prevalence. Figure~\ref{fig:topic-modeling-challenge-categories} shows the distribution of posts by topic and theme.

\subsubsection*{\textbf{Control/Data Flow and Execution}} is the most prevalent challenge category for Ansible practitioners. This category accounts for 10,721 posts, representing 44.4\% of recent discussions.

\subsubsection*{Topic Analysis}
\emph{Loops} is the most represented topic (13.9\%). As one practitioner stated in the posts, ``I don't understand why loops are so hard to do in this language.'' \emph{Loops} are followed by \emph{SSH Connection Issues} (12.7\%), \emph{Variable Management} (12.3\%), \emph{Conditional Execution} (10.6\%), and \emph{Undefined Variables} (10.4\%).

\subsubsection*{Interviews}
Participants described various \textit{control flow issues}. For instance, B1 mentioned frustration with \textit{sequential execution} slowing runs and a \textit{lack of synchronization primitives} (n=2). 
In more advanced setups, complexities grew around \textit{managing dependencies} (n=3).
A handful of participants reflected on \textit{code structure/organization} difficulties (n=5). In contrast, others cited Ansible's inherently \textit{stateless} (n=5) nature as making it simpler to run but also lacking built-in features for preserving or comparing states. The \textit{push model/agentless} (n=5) was both praised for minimal overhead and criticized for complicating continuous enforcement.

\subsubsection*{\textbf{Lack of Support}} is the second most significant theme, accounting for 38.2\% of the recent posts with a total of 9,226 posts.

\subsubsection*{Topic Analysis}

AWX-related issues, where AWX is the open-source upstream project for Ansible Tower, dominated discussions (17.3\%), followed by module execution failures (10.7\%), integration issues (8.5\%), module usage problems (8.2\%), and Windows-related challenges (7.7\%).

\subsubsection*{Interviews}
Interviewees reported missing features that require custom modules (n=3) and limited OS support (n=4). Other participants criticized the \textit{lack of support for certain modules} (n=7). As E4 put it ``There's a lack of support for some modules (...). You often have to write your own or hope someone did it.''

\subsubsection*{\textbf{Reliability}}\label{sec:reliability} emerged as the third most discussed theme, accounting for approximately 37.2\% of all recent posts.

\subsubsection*{Topic Analysis}
Although \emph{Error Handling} is the most represented topic (14.2\%), TopicGPT's interpretation of it is generic. As a result, despite its description being related to the handling of errors in Ansible playbooks, the topic was frequently used in cases where arbitrary errors occurred. Following \emph{Error Handling}, the most discussed issues are related to file management (10.0\%), permissions (7.0\%), JMESPath (5.1\%) and the use of shell commands (4.8\%).

\subsubsection*{Interviews}
Reliability issues primarily emerged around debugging.
Multiple participants reported that Ansible is \textit{hard to debug} (n=10). I7 tries to avoid Ansible to debug complex cases: ``if there's some more complex stages in my playbooks what I try to do is to transfer those to a shell script that I can pass outside of Ansible and I can test those separately.'' B2 described moments when ``it would get stuck for 15 minutes in a supposedly 1 min script before I forced it to close without any visible changes on the terminal.'' Some lamented the lack of built-in \textit{infrastructure debugging tools} (n=5), while others pointed to \textit{bad error messages} (n=5) when modules returned cryptic outputs. As B4 put it, ``I actually believe that there's not a proper way to debugging this type of scripts and tools.''

\subsubsection*{\textbf{Usability}}\label{sec:usability} represents \textasciitilde 18.8\% of recent posts (3,846 posts).

\subsubsection*{Topic Analysis}

Issues related to \emph{Jinja2} are the most prevalent, accounting for 29.1\% of usability-related posts. 
Issues related to quoting (19.7\%), playbook inputs (17.9\%), YAML syntax (9.8\%), and regex (7.2\%) are also prominently discussed. When talking about their issues with YAML, a practitioner mentioned: ``I find YAML or Ansible's interpretation of YAML very frustrating and confusing.''

\subsubsection*{Interviews}
Ansible's \textit{usability} emerged as a significant concern for many participants. 
Participants described Ansible as \textit{challenging for beginners} (n=12), particularly for those \textit{without a coding background} (n=2). 
Others felt they had to make a \textit{mental shift} (n=4), especially if they were used to more imperative scripting. As E8 said, ``If you've never seen Terraform, CloudFormation (...)
it'll look weird and alienating''. Some participants found it \textit{hard to pick the right approach} (n=4). 
Participants also complained about \textit{complex syntax} (n=4). Ansible's reliance on YAML and Jinja introduced the problems of \textit{YAML and Jinja complexity} (n=5 and 6). As E5 said, ``I once spent 8 hours of work figuring out the escaping on a line of SQL %
(...) The hell that is escaping in YAML will never die.''

\subsubsection*{\textbf{Performance and Scalability}} represents approximately 8.1\% of recent posts, with 1,965 posts discussing these issues.

\subsubsection*{Topic Analysis}

Dynamic inventories (29.0\%) and inventory management (27.3\%) dominate discussions, followed by performance (16.3\%), remote execution (7.7\%), and delegation (7.0\%).

\subsubsection*{Interviews}
Participants managing large infrastructures frequently reported \textit{scalability limitations} (n=7).
B1 stated: ``When you run a playbook for like 100 hosts, it gets really confusing trying to make sense of the messages''. 
Participants specifically described problems related to \textit{code awareness} (n=4) in large projects.
A related source of concern was \textit{slow execution} (n=8). While some liked the simplicity of SSH-based workflows, a few individuals cited \textit{agentless} (n=4) as detrimental in high-volume scenarios. E4 was particularly direct: ``It does not scale when you start to manage hundreds of servers, it becomes slow because there is no agent.''

\subsubsection*{\textbf{Maintenance}} challenges appear in 750 posts, which corresponds to about 3.1\% of the total of recent posts.

\subsubsection*{Topic Analysis}
Challenges related to deprecated features and settings in Ansible (32.7\%), dependency management (22.3\%), compatibility issues (18.5\%) (e.g., with legacy OSs), different Ansible versions (13.5\%) or version controlling Ansible projects (11.5\%) are the most relevant topics in the \emph{Maintenance} theme. 

\subsubsection*{Interviews}
Participants noted that \textit{removing configurations is difficult} (n=3) as modifying the playbook alone does not guarantee that older changes get undone. E2 explained that ``Doing things like removing config from hosts tends not to work so well in Ansible,''. \textit{Configuration drift} (n=5) was also mentioned. As I3 described ``Deletes are hard: one cannot merely delete parts of the Ansible playbook and expect those changes to be undone in hosts. Instead, we must change or even add tasks undoing the configuration we want gone.'' Other participants mentioned that Ansible can be \textit{hard to maintain} (n=5) and were concerned about \textit{deprecations} (n=5).

\subsubsection*{\textbf{Information Gaps}} is the least prevalent theme, accounting for only 482 posts, or about 2\% of all recent posts. Of these 482 posts, 480 address the challenges practitioners face with documentation. %

\subsubsection*{Interviews}
Another concern revolved around \textit{information gaps}, particularly incomplete documentation. E2 talked about \textit{misleading information}: ``Another issue is incorrect or misleading answers on forums and newsgroups. Beginners may struggle to distinguish good advice from bad.'' Participants that talked about \textit{documentation gaps} (n=9) described the difficulty of locating relevant or clear information among Ansible's extensive (but sometimes inconsistent) references. Although multiple participants praised Ansible's core documentation (described in~\Cref{sec:RQ2}), they also identified a steep learning curve for interpreting advanced instructions.%

\subsubsection*{\textbf{Topic Evolution}} Figure~\ref{fig:topic-modeling-challenge-categories} shows the yearly
percentage evolution of each topic over the entire Ansible post history until 2024.
Most topics exhibit a stable prevalence over time. However, issues related to AWX, integration and compatibility issues, and documentation increase in prominence, whereas topics such as variable management and module development show a gradual decrease.

\begin{tcolorbox}[
    colframe=orange!20!black,  %
    boxrule=0.8pt,             %
    arc=3pt,                   %
    left=3pt, right=3pt,       %
    top=3pt, bottom=3pt
]
\textbf{RQ1:}
We identified seven core challenge categories: \emph{Control/Data Flow and Execution}, \emph{Reliability}, \emph{Lack of Support}, \emph{Usability}, \emph{Performance and Scalability}, \emph{Maintenance} and \textit{Information Gaps}. Our Q\&A analysis highlighted frequent technical issues (e.g., loops, SSH connectivity), while interviews revealed deeper concerns such as stale configurations, misleading community advice, and complex setup.

\end{tcolorbox}

\section{RQ2: Factors Influencing Ansible Adoption}\label{sec:RQ2}
This section analyzes how interview participants first encountered Ansible and the reasons they selected it over other IaC solutions. %

\subsubsection*{Accessibility \& Usability.}
A significant theme in our \textit{Advantages} codebook is the role of Ansible's accessibility in driving adoption. Several participants described it as \textit{Easy to use} (n=13),  \textit{Easy to onboard to new people} (n=7), and with a \textit{Low Barrier to Entry for Setup} (n=6) highlighting how they only needed to master a limited set of concepts—such as basic playbook structures and inventory configuration—before managing a wide range of environments. B1 remarked, ``To get things done, I only really needed to understand a small set of concepts regarding how Ansible works''. Participants also cited Ansible's \textit{Readable syntax} (n=6),
E6 highlighted that ``[we] were looking into Ansible because it has a nice YAML syntax''.

\subsubsection*{Tool Qualities \& Ecosystem.}
Beyond accessibility, Ansible was praised for its \emph{tool qualities}, with participants discussing its \textit{Scalability} (n=5).
Another advantage of Ansible is its \textit{Flexibility}, since it can be a \textit{One-Tool Solution} (n=2) capable of managing heterogeneous environments (e.g., Windows, Linux, or network devices). As E3 puts it, the ``One tool that does it all''. 
Several participants emphasized that Ansible's \textit{Compatibility} (n=6) with various operational requirements and existing workflows set it apart from other tools, and its \textit{Agentless} (n=9) model that depends solely on SSH. %
\textit{Idempotence} (n=2) was also considered important for participants, ensuring the same playbook can be applied repeatedly to converge infrastructure into a consistent state. As I3 described: ``Some commands are dangerous to invoke twice.''
From an \textit{Ecosystem} standpoint, respondents highlighted features such as \textit{Easy to extend} (n=4), and the fact that \textit{It's implemented in Python} (n=2). 

Participants frequently framed their rationale in comparative terms,
highlighting Ansible's advantages over alternatives like Puppet, Chef, and Salt. 
A common comparative advantage cited was the elimination of agent management overhead.

E5 provided an illustrative anecdote highlighting Ansible's comparative operational model advantages:
``(...) a place I was working was using Puppet, and the admin was woken up at 3 am from a page, fixed the issue, went to bed, and then Puppet restored the broken configuration, so he woke up again, fixed the Puppet, and then went to bed. Then, was woken up again because Puppet updated itself from git.''

\subsubsection*{Community.}
 A third central theme centers on Ansible's Community. The tool's \textit{Popularity} (n=4) and \textit{Community}  (n=4) are significant advantages. E1 noted: ``Ansible's advantages (...) %
 are 
 it being a well-supported IaC tool. Since there is a large community around it, it is easier to find solutions to specific problems online.'' I1 summarized it as: ``Ansible was just the hot topic at the time.''
Participants also mentioned that Ansible's strong market presence often led their organizations to mandate its use (n=8) (``it was standard by the time I joined the team'', I3).
Moreover, if it was not mandatory, it was often \textit{Recommended} (n=3) by colleagues or professors. %

\begin{tcolorbox}[
    colframe=orange!20!black,  %
    boxrule=0.8pt,             %
    arc=3pt,                   %
    left=3pt, right=3pt,       %
    top=3pt, bottom=3pt,
    breakable,
    enhanced
]
\textbf{RQ2:}
 Ansible's adoption is influenced by its ease of use, low barrier to entry, and agentless design. Participants highlighted that minimal setup requirements and an approachable syntax enabled them to quickly manage diverse infrastructures. They also praised Ansible's extensibility,  strong community presence, and popularity.
\end{tcolorbox}

\section{RQ3: Improving Ansible}\label{sec:RQ3}
While participants generally acknowledged Ansible's strengths, they also offered concrete \textit{recommendations} to improve.

\subsubsection*{Documentation and Learning Resources.} A recurring concern expressed by participants (n=10) was the need to \textit{improve documentation}. E4 described the current official documentation as scattered, noting that it ``is poor considering the other configuration tools (take a look at Puppet's). The websites are not appealing and not properly organized''.
I3 identified specific documentation challenges regarding module organization: ``Documentation needs a big revamp: it's not clear what modules are built-in or part of Ansible Galaxy.'' E8 found the documentation helpful but noted that ``sometimes they don't [provide] all the examples''
The participants also called for \textit{better onboarding} resources (n=6) and \textit{more education} (n=9). As E3 put it ``When you educate beginners, they can identify incorrect information and avoid it.''

\subsubsection*{Debugging and Troubleshooting.} Several participants highlighted limitations in Ansible's current debugging capabilities. \textit{Better debugging tools} (n=5) and \textit{improved error messages} (n=6) were frequently mentioned. E2 stressed that ``error messages when plugins fail could be improved,'' as they can sometimes be ``misleading''. I2 reinforced this point, saying that ``more specific error messages would be very helpful. Understanding exactly where an issue presents is the most important part of the debugging process''.
Another concept frequently raised was the lack of \textit{rollback-on-failure} support (n=5).

\subsubsection*{Performance and Scalability.} Participants also critiqued performance and mentioned the need to \textit{improve speed} (n=7) and improve \textit{parallelization} (n=4). As E2 put ``Slow runs is an eternal pain point.'' E1 advocated for integrating solutions akin to Mitogen~\cite{mitogen_2025}, to help optimize remote execution over SSH to reduce overhead and B1 proposed a tag-based approach: ``Having the ability to set group tags for tasks, and have groups of tasks be able to run in parallel''.%

\subsubsection*{Advanced Programming Capabilities.} A group of participants called for the expansion of Ansible's programming model. While some participants advocated for \textit{more high-level language concepts} (n=3) (e.g., real loops, and complex conditionals), others (n=5) stressed \textit{state management}. %
E7, for instance, argued for ``functionality in Ansible to handle state management effectively'' and provided an extensive analysis of Ansible's state management limitations, saying, ``even though Ansible claims to be completely stateless, which is supposed to be its strength, I'm not so convinced. Without state management, it's challenging to handle certain problems.''

\begin{tcolorbox}[
    colframe=orange!20!black,  %
    boxrule=0.8pt,             %
    arc=3pt,                   %
    left=3pt, right=3pt,       %
    top=3pt, bottom=3pt,
    breakable,
    enhanced
]
\textbf{RQ3:}
Interviewees emphasized the need for better documentation, improved debugging, and enhanced error reporting. Many highlighted performance optimizations, including faster execution and parallelization, while others advocated for rollback support and stronger third-party integrations.
\end{tcolorbox}

\section{Discussion}
\label{sec:discussion}
This section discusses the implications of our findings for Ansible as well as for other IaC technologies.

\subsection{Failure Locality Guarantees}

Our analysis indicates that many Ansible challenges stem from difficulties in localizing failures during playbook execution.
High-frequency topics such as \emph{Error Handling}, \emph{Undefined Variables}, and \emph{YAML, Jinja and Quoting-related syntax issues} frequently involve failures that surface far from their root causes. Interview participants consistently described situations where Ansible ``gets stuck'', produces misleading errors, or provides insufficient context to determine where execution went wrong. Several interviewees explicitly noted limitations in the current error messages and expressed a need for more informative feedback. Notably, these observations were made by interviewees with substantial Ansible experience (on average, approximately 4 years), suggesting that these difficulties persist beyond initial onboarding.

We conceptualize this recurring pattern as \emph{failure locality}: the degree to which an IaC tool enables practitioners to identify the origin and propagation of failures during execution. In Ansible, low failure locality manifests when errors arise late, lack actionable context, or are disconnected from their causes.

This lack of failure locality is reflected in the interviews, where half of the participants reported that Ansible is hard to debug and frequently cited the absence of meaningful progress or termination feedback during execution. As discussed in Section~\ref{sec:reliability}, a participant reported a playbook that stalled for 15 minutes despite being expected to run in one minute, with no visible terminal feedback.
In such cases, users cannot determine whether the execution is progressing or identify the task responsible for the stall.
As a result, failure localization often requires manual interruption, additional logging, or trial-and-error instrumentation.

This lack of execution feedback is also reflected in community discussions. We identified 160 recent posts ($\approx$0.7\%) whose titles contain terms such as \emph{stuck}, \emph{hang}, or \emph{freeze}, suggesting that non-terminating or stalled executions are a recurring concern. We restricted the search to post titles to avoid overcounting incidental mentions and to focus on posts where non-termination was the central issue. These reports align with documented Ansible issues that describe stalled executions with limited diagnostic output~\cite{ansible_issue_18305,ansible_issue_30411}.
Many of these posts were attributed topics related to SSH connection issues (50 posts), AWX (23), and WinRM (10), 
suggesting that stalled executions often arise at integration boundaries, such as remote connections and execution orchestration, where Ansible provides limited visibility into progress and blocking conditions.

\begin{tcolorbox}[
  action_style,
  breakable,
  enhanced
]
\textbf{Insight for Ansible:}  
Ansible could provide phase-aware execution progress and stall diagnostics at integration boundaries such as remote connections. 
Surfacing the current execution phase (e.g., connection establishment or module execution), along with task and host context, and warning when a phase stalls would help practitioners localize blocking behavior without manual interruption.
\end{tcolorbox}

The low failure locality in Ansible is not limited to a single class of failures. Our analysis shows that failures related to YAML syntax and quoting exhibit particularly poor locality characteristics, making them among the most prominent usability challenges faced by practitioners. 
An inspection of the Q\&A posts revealed a recurring error message that appears in 706 posts: \emph{``The error appears to have been in file X: line Z, column Y, but may be elsewhere in the file depending on the exact syntax problem.''}

Such messages acknowledge the uncertainty about the fault's location, forcing users to search beyond the reported location. This behavior undermines failure locality, as the system is unable to confine the failure to the construct that caused it. Instead of narrowing the search space, the diagnostic output expands it, increasing the cognitive effort required to diagnose even minor errors.

The interview data further illustrate the practical impact of this loss of locality. As mentioned in Section~\ref{sec:usability}, a \emph{participant with 9 years of Ansible experience} spent several hours debugging a YAML quoting issue. This example highlights how small, localized mistakes can propagate into failures that are difficult to attribute, diagnose, and correct due to ambiguous or misleading error reporting.

\begin{tcolorbox}[
  action_style,
  breakable,
  enhanced
]
\textbf{Insight for YAML-based IaC technologies:} 
Improving debugging effectiveness requires compensating for YAML's weak syntactic failure locality. 
YAML-based IaC tools could provide early, schema-guided validation, precise diagnostics for quoting and escaping errors that identify the exact responsible construct instead of approximate file locations, and clear distinctions between parsing, templating, and execution failures, thereby reducing the diagnostic search space.
\end{tcolorbox}

\subsection{Blurred Language Boundaries}

Our Q\&A analysis indicates that issues related to \emph{Loops}, \emph{Jinja2}, \emph{Conditional Execution}, and \emph{Undefined Variables} are among the most frequently reported by practitioners
Moreover, these topics exhibit strong co-occurrence in Ansible-related discussions. Table~\ref{tab:topic_cooccurrences}  reports the pairwise co-occurrences and coverage among these four topics. Additionally, across all topics, roughly one-fifth of posts associated with any topic also involve at least one of the others.
These results suggest that practitioners reason about these concepts together. They experience them as a tightly coupled cluster of challenges that emerge at the intersection of Ansible's execution model and the Jinja2 templating system. This is consistent with our interview data, in which participants repeatedly highlighted difficulties with Jinja2, mainly experienced practitioners with an average of seven years of Ansible experience, and suggested the addition of higher-level language constructs as a potential improvement.

\begin{table}[]
\centering
\caption{Pairwise co-occurrences between topics \emph{Jinja2}, \emph{Loops},
\emph{Conditional Execution}, and \emph{Undefined Variables}. Each cell shows the count and the coverage of the row topic (in \%).}
\footnotesize
\label{tab:topic_cooccurrences}
\begin{tabular}{lcccc}
\toprule
 & \textbf{Jinja2} & \textbf{Loops} & \textbf{Cond. Exec.} & \textbf{Undef. Vars.} \\
\midrule
\textbf{Jinja2} 
 & -- 
 & 134 (10.1\%) 
 & 67 (5.9\%) 
 & 95 (7.2\%) \\

\textbf{Loops} 
 & 134 (9.0\%) 
 & -- 
 & 129 (8.7\%) 
 & 76 (5.1\%) \\

\textbf{Cond. Exec.} 
 & 67 (5.1\%) 
 & 129 (11.4\%) 
 & -- 
 & 79 (7.0\%) \\

\textbf{Undef. Vars.} 
 & 95 (8.5\%) 
 & 76 (6.8\%) 
 & 79 (7.1\%) 
 & -- \\
\bottomrule
\end{tabular}
\end{table}

Playbooks operate at the intersection of multiple languages and abstraction layers: YAML as a data serialization format, Ansible-specific schema extensions (e.g. when), and Jinja2 as an embedded templating language. 
Their composition creates blurred semantic boundaries that are difficult for practitioners to internalize. 

This interaction helps to explain an apparent tension in our findings. In RQ2, participants frequently described Ansible's readable syntax as a key factor influencing adoption. However, our results indicate that this readability may not scale uniformly with the complexity of playbooks. As playbooks become increasingly entangled across language boundaries, practitioners must simultaneously determine in which language they are currently writing in, which evaluation model applies, and where expressions are interpreted. Maintaining an accurate mental model of these shifting boundaries imposes considerable cognitive overhead.

\subsubsection*{Nested Loops}
A manifestation of this boundary breakdown appears in the handling of nested loops, which account for 7.3\% of all loop-related discussions.
Notably, Ansible's official documentation explicitly discourages nested loops~\cite{ansible_loops_2026}: ``The simplest way to ‘nest' loops is to avoid nesting loops, just format the data to achieve the same result''. Instead, Ansible recommends that users restructure data and rely on Jinja2 filters, to simulate nested iteration. This design choice introduces a tension between Ansible's declarative model and the imperative nature of iteration. While loops are exposed as imperative constructs, their expressiveness is deliberately constrained, pushing users toward a functional programming style implemented via Jinja2 filters. As a result, users are expected to translate imperative intent into functional transformations, increasing cognitive load and reducing readability.

\subsubsection*{Ansible-specific filters}
This tension is further amplified by Ansible's reliance on Ansible-specific Jinja2 extensions and filters. For example, the documentation prescribes the use of the \emph{dict2items} filter, an Ansible-specific addition, to iterate over dictionaries~\cite{ansible_iterating_dict_2026}. 
We obtained the list of Ansible-specific filters from the official documentation~\cite{ansible_builtin_collection_2026}
 and subsequently searched the Q\&A posts for corresponding invocation patterns for each filter.
 We identified the usage of Ansible-specific filters in user-provided examples in 3,014 posts, corresponding to 12.5\% of all recent posts.

We then assessed whether these filters were central to the issues discussed in the posts. To this end, we selected posts that not only matched the aforementioned pattern but also referenced the same filter, or a semantically related term, in the post title. For example, for the \emph{regex\_findall} filter, we additionally searched for the terms \emph{regex}, \emph{regular}, \emph{expression}, and \emph{findall}. Of the 3,014 posts containing filter usages, 706 satisfied this criterion, indicating that at least 3\% of all recent posts are related to the use of Ansible-specific filters.

Although technically effective, such constructs deepen the entanglement between Ansible and Jinja2, forcing users to learn not only the Jinja2 semantics but also the customized Ansible dialect. Rather than separating concerns, this approach layers declarative configuration, imperative execution, and functional transformation into a single syntactic space, making it harder for users to predict behavior and diagnose errors.

\subsubsection*{Templating boundaries} Inconsistencies in templating boundaries further obscure the mental model practitioners must maintain. Conditional execution (\emph{when}) implicitly evaluates Jinja2 expressions without requiring the explicit Jinja2 delimiters used in Ansible, while loops require Jinja2 expressions to be explicitly enclosed within these delimiters. In fact, on April 14, 2025, a commit~\cite{ansible_templating_deprecation_commit} was pushed to Ansible that introduces a new deprecation warning that \emph{``Conditionals should not be surrounded by templating delimiters such as \{\{ \}\} or \{\% \%\}. This feature will be removed from ansible-core version 2.23''}. A similar warning had already been introduced in Ansible 2.3, as one collected post mentions~\cite{ansible_warning_23_forum}.
The recurrence of this guidance across versions suggests a long-standing source of confusion. Consistent with this, we observe that this warning message appears 37 times in recent posts.
Such inconsistencies make it unclear when users are writing Jinja2, how expressions are parsed and evaluated, and consequently weaken boundary visibility.

Taken together, these findings suggest that many of Ansible's usability challenges may not be due to individual features, but to boundary friction between languages and paradigms that overlap.

\begin{tcolorbox}[
  action_style,
  breakable,
  enhanced
]
\textbf{Insight for Ansible:} 
Our findings suggest the need to revisit the design and exposure of control-flow abstractions, with a focus on making language boundaries explicit and reducing cross-paradigm leakage. Ansible could benefit from first-class expressive control-flow constructs that align with the mental models of practitioners, including native support for nested iteration and richer conditionals, thus reducing the dependency on Jinja2 as a workaround language and lowering the cognitive overhead introduced by functional-style filters.\\

In parallel, Ansible could strive for consistent templating semantics by standardizing expression evaluation rules across constructs or explicitly signaling evaluation contexts. Treating Jinja2 as a clearly delineated component, with explicit role, scope, and limitations in both syntax and documentation, would improve learnability and reduce boundary confusion.
\end{tcolorbox}

\begin{tcolorbox}[
  action_style,
  breakable,
  enhanced
]
\textbf{Insight for designers of IaC technologies:} 
Our results highlight the risks of polyglot configuration models that blend declarative, imperative, and templating paradigms without clear separation. 
Our findings show that boundary ambiguity can become a primary source of user difficulty.
Future IaC tools could therefore treat boundary design as a first-class concern, explicitly modeling where data declaration ends, where control flow begins, and how (or whether) templating is allowed to intervene.
\end{tcolorbox}

\subsection{Documentation Gaps}

Documentation is the primary means by which users learn and troubleshoot IaC tools, however, our findings show that Ansible's documentation exhibits both explicit gaps (directly reported as missing or inadequate) and implicit gaps, inferred from recurring questions that existing documentation fails to resolve.

Topic modeling indicates that documentation issues constitute a subset of the \emph{Information Gaps} theme (480 posts, ~2\% of recent posts).  
Participants across experience levels (n=10) repeatedly called to \textit{improve documentation}. E4 described the current official documentation as scattered, noting that ``strong concepts are scattered over different pages.'' 
E8 found the documentation examples helpful but noted that ``sometimes they don't [provide] all the examples''.

Because topic modeling captures only explicit discussion of documentation, we conducted a post-hoc analysis to identify documentation issues embedded in technical posts. This analysis shows that 5.4\% of all posts (1,301) report documentation problems (2.5× higher than topic modeling suggests). This result comes from a two-stage process: a keyword search followed by manual validation. The search identified 1,858 posts (7.7\%) mentioning ``documentation'' or ``docs'', which were then refined through review to isolate genuine documentation issues.
Using a conservative proportion of 0.5 with finite-population correction, we sampled 353 posts to achieve a 95\% CI with a $\pm 5\%$ error margin. One author coded the posts, with a second author independently reviewing the classifications as either true positives (e.g., missing, unclear, or mismatched information) or false positives. Of the sample, 247 posts (70.0\% $\pm$ 5\%, 95\% CI) were true positives and 106 (30.0\% $\pm$ 5\%, 95\% CI) false positives, thus, about $5.05\%$ of recent posts genuinely report documentation issues.
We also saw that the percentage of posts reporting documentation problems has increased over time from approximately 4.2\% in 2016-2020 to 7.6\% in 2024 and 10.9\% in early 2025. Because this post-hoc analysis relies on manual interpretation of discussion content, the resulting proportion should be interpreted as an approximate estimate rather than an exact prevalence.
Documentation concerns also arise while discussing other technical topics, including Module Usage ($\sim$80 posts), Integration Issues ($\sim$74), Python Issues ($\sim$97), and Error Handling ($\sim$53).
Users in posts report missing documentation (``Couldn't find Ansible documentation on GPG''~\cite{ansible_gpg_missing_docs_2019}), mismatches between documentation and behaviour (``If I read the documentation to the letter it should fail''~\cite{ansible_loops_conditionals_doc_mismatch_2024}), and finding existing documentation insufficient (``the documentation is still very vague about how this actually works and behaves''~\cite{ansible_failed_successful_reddit_2023}).

\begin{tcolorbox}[
  action_style,
  breakable,
  enhanced
]
\textbf{Insight for IaC tools:}
The finding that 5.4\% of recent posts report documentation issues within technical discussions suggests that documentation gaps surface during implementation rather than through explicit documentation queries. Although high-quality documentation lowers barriers to entry~\cite{robillard2009makes, aghajani2019software}, many tools fail to keep pace with rapid feature growth and expanding user bases~\cite{aghajani2020software}. Tool developers can apply our topic modeling approach to mine community Q\&A data and prioritize documentation improvements by technical domain, focusing on topics that most frequently co-occur with documentation mentions. While demonstrated on Ansible, this method generalizes to other IaC and software tools. %
\end{tcolorbox}

\subsection{Execution Backends and Scalability}

Performance and scalability concerns are reflected in both our quantitative and qualitative results. Performance and Scalability account for 8.1\% of recent posts, and within this category, posts explicitly classified under \emph{Performance} represent 16.3\%. An analysis of topic co-occurrence within \emph{Performance} posts shows that \emph{AWX} (10.6\%), \emph{Loops} (10.2\%), \emph{SSH Connection Issues} (9.0\%), \emph{Error Handling} (5.6\%), and \emph{Asynchronous Execution} (5.0\%) are the most frequent co-occurring topics, where percentages denote the proportion of \emph{Performance} posts in which each topic appears. The prominence of \emph{SSH Connection Issues} and \emph{Asynchronous Execution} indicates that performance problems are frequently associated with remote execution overhead and limitations in parallel execution models.

Interview participants emphasized \textit{slow execution} and the need to \textit{improve speed} and \textit{parallelization}, particularly when managing large infrastructures. Several participants identified the \textit{agentless} execution model as a contributing factor to scalability limitations, with one E2 describing slow runs as ``an eternal pain point''. 

This concern must be viewed in light of the factors driving Ansible's adoption. Participants highlighted the agentless, SSH-based execution model as a key advantage. Rather than suggesting that agentless execution should be abandoned, our findings point to the need for execution backends that preserve this model while mitigating its performance costs.
In this context, an experienced participant advocated for integrating solutions similar to Mitogen~\cite{mitogen_2025} to reduce SSH overhead and improve execution efficiency. 

Mitogen is an alternative execution framework for Ansible that optimizes remote task execution by minimizing repeated SSH connection setup and Python interpreter initialization. Instead of spawning a new remote process for each task invocation, Mitogen establishes persistent execution contexts and reuses them across tasks, significantly reducing communication and startup overhead,
while preserving Ansible's agentless execution model.

\begin{tcolorbox}[
  action_style,
  breakable,
  enhanced
]
\textbf{Insight for Ansible:}
Mitogen illustrates that Ansible's performance limitations are not inherent to declarative IaC workflows, but are %
shaped by the characteristics of its execution backend. An implication for Ansible is therefore to treat optimized execution frameworks, such as Mitogen-like approaches, as first-class supported components with clear integration into the platform.
Such support would allow users to systematically address \textit{slow execution} and \textit{parallelization} challenges in large-scale deployments, rather than relying on external or ad hoc optimizations.
\end{tcolorbox}

\subsection{Future Work}
We identify opportunities for Ansible's evolution and broader design trade-offs in IaC.
Our concept of failure locality warrants engineering validation through the evaluation of phase-aware diagnostics and error reporting in controlled experiments.
The blurred language boundaries suggest the need for design guidelines and patterns for composing declarative, imperative, and templating paradigms in IaC languages, potentially via prototype DSL development.
Mining large-scale Ansible repositories could quantify the prevalence of identified anti-patterns (e.g., nested loops,  Jinja2 complexity) and enable correlation with code quality metrics.
We also argue that our mixed methodology of combining topic modeling of community discussions with targeted practitioner interviews to identify challenges is replicable in other contexts. Applying this methodology to tools such as Terraform or Kubernetes could enable a cross-tool comparison of challenge patterns and inform tool selection and improvement priorities across the IaC ecosystem.

\section{Threats to Validity}

\subsubsection*{Threats to Internal Validity.} The use of LLMs via TopicGPT introduces a risk of hallucination and unreliable outputs, potentially affecting result precision. We mitigated this through a two-stage manual validation during topic generation and topic assignment. Semi-structured interviews also risk interviewer bias; this was mitigated by using a consistent protocol and piloting the interviews.

\subsubsection*{Threats to External Validity.} Our findings may be limited by practitioners discussing Ansible on Q\&A platforms not included in our dataset; we mitigated this by collecting data from three distinct platforms. While some findings may generalize to other IaC tools, particularly configuration management tools, further research is needed to confirm their applicability. Interview participants were primarily recruited online, introducing potential self-selection bias and limiting representativeness across expertise levels. As with qualitative research generally, interview results provide contextual insights but are not statistically representative of the broader Ansible community.
Our analysis relies on public Q\&A platforms whose usage patterns are changing as generative AI adoption increases~\cite{burtch2024consequences}. We mitigated this by restricting analysis to posts from the five years prior to data collection; the resulting corpus of 24,132 posts still exhibits diverse topics, indicating continued platform relevance. Nevertheless, as generative AI increasingly addresses common issues, future forum discussions may shift toward more complex problems, potentially affecting long-term generalizability.

\section{Related Work}
To identify challenges that practitioners face in IaC, Rahman et al. analyzed 2,758 Puppet-related questions from Stack Overflow~\cite{rahman2018questions}, identifying 16 major categories, many of which (e.g., \emph{Syntax Error}, \emph{Filesystem}) also appear in our study. While their focus is Puppet, we study Ansible, a more widely used configuration management tool as of 2024, and consider more recent posts and interviews.
Furthermore, Rahman et al. analyzed questions posted between January 2010 and December 2016, while our interviews and posts reflect the scenario as of 2024.

Guerriero et al. conducted 44 semi-structured interviews with practitioners to examine IaC practices, tools, and challenges~\cite{guerriero2019adoption}, highlighting issues such as testing and debugging. Our study provides a more granular analysis and identifies Ansible-specific issues such as lack of failure locality and blurred language boundaries.

Begoug et al. collected 52,692 questions and 64,078 responses from Stack Overflow to investigate challenges developers encounter with IaC \cite{begoug2023infrastructure}. The dataset includes questions related to Ansible, Pulumi, Terraform, CloudFormation, and other IaC technologies. The authors grouped questions into topics using the Latent Dirichlet Allocation (LDA) method optimized with a Genetic Algorithm (GA) for parameter fine-tuning.
Our analysis of practitioners' questions follows a similar approach. However, we collected data from two additional sources: Reddit and the Ansible forum. Additionally, instead of using LDA, we use TopicGPT, which not only provides interpretable topics but also produces topics that align better with human categorizations~\cite{pham2024topicgpt}. 
We both found common topics such as \emph{File Management} and \emph{Templating}, but we found additional challenges faced by practitioners such as Jinja and AWX issues.
Another key distinction between our study and those described above is that while they focus either on interviews or on questions posted by practitioners, our study employs a mixed-method approach that integrates data from both sources.

To our knowledge, the only other mixed-methods study on IaC challenges is that by Tanzil et al. \cite{tanzil2023mixed}. However, their study addresses DevOps challenges in general, with IaC as just one category. They used LDA to perform topic modeling on over 174,000 Stack Overflow posts, identifying 23 distinct topics. These were grouped into four categories, with IaC being one of them, encompassing seven specific subtopics. Finally, the authors validated and extended their findings with a survey of 21 professional DevOps practitioners. 
Although there is some overlap in the findings between our study and that of Tanzil et al. (e.g.,  both identify challenges related to \emph{File Management} and \emph{Syntax}), 
we focus exclusively on challenges related to Ansible and IaC, which provides a more focused scope with different insights such as difficulties with YAML and the risks of designing a polyglot technology. We also employ a different topic modeling approach and focus on interviews over surveys.

\section{Conclusion}
We combined a large-scale quantitative analysis of 59,157 online Ansible discussions with 20 practitioner interviews to identify key challenges, contextualize their real-world impact, and propose solutions. We found issues related to control flow, performance bottlenecks, limited debugging capabilities, and inconsistent documentation, highlighting the tensions among usability, flexibility, and scalability in Ansible's ecosystem. 
\newc{}{We recommend the use of phase-aware stall diagnostics, YAML validation, better control-flow constructs, optimized execution backends, and topic-guided documentation prioritization.} These contributions, alongside broader reflections on community-driven knowledge, provide a replicable roadmap for assessing and improving IaC solutions.

\begin{acks}
We
would like to thank Claúdia Mamede, who helped us design our main diagram.
This work was supported by national funds through 
Fundação para a Ciência e a Tecnologia, I.P. (FCT) under grants 
PRT/BD/153739/2021 and BD/04736/2023, and projects 
UID/50021/2025 (DOI: \href{https://doi.org/10.54499/UID/50021/2025}{10.54499/UID/50021/2025}), UID/PRR/50021\-/2025 (DOI: \href{https://doi.org/10.54499/UID/PRR/50021/2025}{UID/PRR/50021/2025}), and by 
the InfraGov project, with ref. n. 2024.07411.IACDC (DOI: 
\href{https://doi.org/10.54499/2024.07411.IACDC}{10.54499/2024.07411.IACDC}), funded by the `Plano de Recuperação e Resiliência (PRR)' under the investment `RE-C05-i08 - Ciência Mais Digital', measure `RE-C05-i08.M04' (in accordance with the FCT Notice No. 04/C05 i08/2024), framed within the financing agreement signed between the `Estrutura de Missão Recuperar Portugal (EMRP)' and the FCT as an intermediary beneficiary.
It was also co-funded by the European Union through the Lisboa 2030 Programme (ERDF) and by national funds through FCT, I.P., under project no. 15018 (SafeIaC project, DOI: \href{https://doi.org/10.54499/2023.18089.ICDT}{10.54499/2023.18089.ICDT}).

\end{acks}

\bibliographystyle{ACM-Reference-Format}
\bibliography{references,dejavu}

\appendix

\end{document}